\documentstyle[aps,prl,psfig,floats]{revtex}
\begin{document}
\draft
\twocolumn[\hsize\textwidth\columnwidth\hsize\csname@twocolumnfalse\endcsname

\title{The effects of interface morphology on Schottky barrier heights: \\
a case study on Al/GaAs(001)}

\author{Alice Ruini,$^{1,2}$ Raffaele Resta,$^{1,3}$ and 
Stefano Baroni$^{1,2,4}$}

\address{$^1$INFM -- Istituto Nazionale di Fisica della Materia \\
$^2$SISSA -- Scuola Internazionale Superiore di Stud\^\i\ Avanzati, Via
Beirut 4, 34014 Trieste, Italy \\ $^3$Dipartimento di Fisica Teorica,
Universit\`a di Trieste, Strada Costiera 11, 34014 Trieste, Italy \\
$^4$CECAM -- Centre Europ\'een de Calcul Atomique et Mol\'eculaire, 46 All\'ee
d'Italie, 69007 Lyon, France}

\date{April 1997}

\maketitle

\begin{abstract} The problem of Fermi-level pinning at
semiconductor-metal contacts is readdressed starting from
first-principles calculations for Al/GaAs. We give quantitative
evidence that the Schottky barrier height is very little affected by
any structural distortions on the metal side---including elongations of
the metal-semiconductor bond (i.e. interface strain)---whereas it
strongly depends on the interface structure on the semiconductor side.
A rationale for these findings is given in terms of the interface
dipole generated by the ionic effective charges. \\ 
\end{abstract}

]

\narrowtext
   
Despite several decades of extensive experimental and theoretical
work,\cite{Monch} the key factors affecting the Fermi-level pinning at
metal-semiconductor contacts have not yet been clearly assessed. Since the
microscopic morphology of the interface is not experimentally accessible,
the controversy concerns even the very basic issue as to whether the
pinning is determined by {\it intrinsic} interface states which exist even
at an abrupt ideal interface, or by {\it extrinsic} electronic states
arising from native defects. The experimental data are of little help in
discriminating between different theoretical pictures, given that the
value of the barrier for a given semiconductor depends very little on the
nature of the metal,\cite{Monch} and for a given metal/semiconductor pair
it depends even less on the direction of growth. {\it Ab initio}
calculations---though necessarily limited to rather idealized situations
and affected by basic approximations necessary to cope with the complexity
of the electronic many-body problem---allow instead, by their very nature,
to have full control on the way the details at the atomic scale of a given
system affect the various physical properties under investigation. In this
sense, {\it ab initio} calculations are complementary to the experimental
investigations and, in the specific case of metal-semiconductor contacts,
they have in fact provided in recent year a great deal of quantititave
information that any successful model will have to account for. 
\cite{vanSchilfgaarde90,Dandrea93,vanSchilfgaarde94,Needs94,Bardi96,Berthot96,Peressi97} Because of this theoretical work, the following facts are now well
established: the barrier height {\it does} depend of the nature of the
metal;\cite{vanSchilfgaarde90} it {\it does} depend on the
crystallographic direction; and furthermore for a given crystallographic
direction of growth it even depends on the microscopic morphology of the
interface.\cite{Needs94} The electronic mechanisms governing the value of
the Schottky barrier---as well as their variations as a function of the
microscopic morphology of the interface---have not been systematically
investigated so far and are basically unknown. Here we provide a
contribution in this direction, by studying the barrier-height variations
induced in Al/GaAs(001) by several structural and morphological
perturbations which are switched on and off in our computational
framework. Our calculations provide a microscopic probe for the nature of
the interface---including its ``effective'' thickness---and for the
electronic response phenomena responsible for the barrier height. The
crucial role of the effective dynamical charges of interface ions is
elucidated.

The Al/GaAs(001) interface is sp-bonded and almost perfectly
lattice-matched (1\% mismatch); because of the actual growth
conditions, the semiconductor is likely to be As-terminated. At
variance with previous first-principles work, we don't aim at a
detailed modeling of the the real interface; instead, we concentrate
on a reference system as simple as possible, so as to evidentiate the
leading effects induced by controlled variations of the interface
morphology. We assume therefore a defect-free epitaxial geometry as a
working hypothesis. On the same ground, we study here an ideal,
unstrained, interface where the metal is a ``fake'' Al, perfectly
lattice-matched to GaAs, and hence retaining its cubic structure in
the epitaxial overlayer. Strain effects, although quantitatively
sizeable, are considered spurious in the present analysis (see however
some considerations below).

\begin{figure}  \centerline{\psfig{file=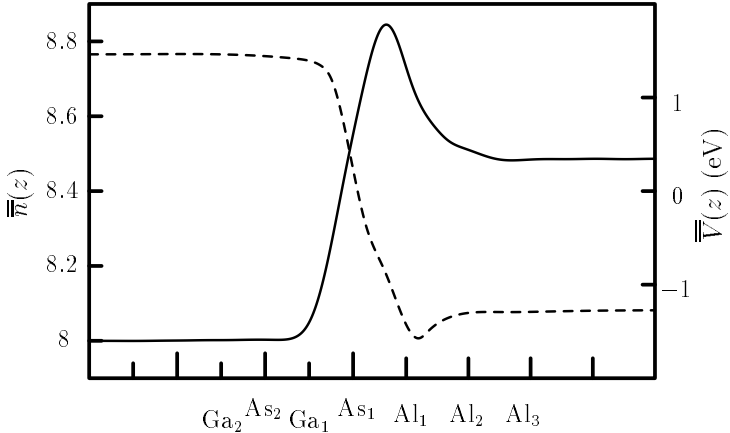,height=7cm}}
\caption[1]{One half of the 31-atoms computational supercell
modeling the (001) interface: in abscissa we have the $z$ coordinate normal
to the interface. The positions of the atomic layers are shown using
vertical bars of different length: the labels identify the layers closest to
the junction.  The functions displayed are the macroscopic
averages---defined as in Ref.  8---of the electronic density (solid line,
scale at the left), and of the total electrostatic potential (dashed line,
scale at the right). } \label{base} \end{figure}

The interfaces are modeled with periodically repeated supercells. The
results for our (001) interface are obtained with a supercell where
the semiconductor slab is chosen with double As-termination, thus
containing two equivalent junctions. In this geometry the metal and
the semiconductor cubic axes are rotated by 45 degrees around the
growth direction, and the lattice-matching condition sets the ratio of
the two cubic lattice constants equal to $1/\sqrt{2}$. A typical
supercell, such as for the calculation shown in Fig.~\ref{base},
contains 9 Al layers, 6 Ga layers, and 7 As layers, for a total of 31
atoms (there are two Al atoms per layer). We focus on the barrier
between the GaAs valence-band edge and the Al Fermi level, relevant
for hole carriers and hence indicated as $\Phi_p$. As
usual,\cite{Pechino92} the barrier height can be partitioned into two
contributions: the electrostatic {\it potential lineup} across the
interface $\Delta V$, and the {\it band-structure term} $\Delta
E_p$. The latter is the difference between the Fermi energy of the
metal and the valence-band edge of the semiconductor, each measured
with respect to the average of the electrostatic potential of the
corresponding crystal. The potential lineup is an interface-specific
property and must be extracted therefore from supercell calculations,
while the band-structure term is the difference between bulk
properties of the two constituents, and hence it is obtained from
independent calculations for crystalline GaAs and Al. The
calculations have been performed using density-functional theory in
the local-density approximation, using pseudopotentials and plane
waves;\cite{pseudo} reciprocal space integrations are performed on a
Monkhorst-Pack special-point grid,\cite{Monkhorst76} using the
smearing technique of Ref.~\onlinecite{Methfessel89} (see also
Ref.~\onlinecite{deGironcoli95}, with which we share several technical
ingredients). The 31-atom supercell calculations are well converged
using a (10,10,2) grid and a smearing parameter $\sigma\!=\!0.01$ Ry.

The relevant results extracted from a typical 31-atom supercell are
shown in Fig.~\ref{base}. The solid line is the macroscopic
average\cite{Pechino92} of the electron density, in units of (valence)
electrons per semiconductor cell. In these units the bulk density of
the semiconductor is 8, whereas the one of the (fake) bulk Al reaches
the value of 8.485, which in fact equals $6$ (number of electrons 
in one periodicity of Al) times $\sqrt{2}$ (ratio between the periodicity
in the GaAs region and the one in the Al region).   
Because of symmetry, we show only one half of the supercell. It
is easily realized that the actual density reaches its bulk value very
close to the junction, thus showing that the supercell is large enough
to model the isolated (and neutral) interface. Solution of the Poisson
equation for the {\it total} charge (electronic and ionic) yields the
macroscopic average of the electrostatic potential, shown in the same
figure as a dashed line. The lineup between the plateaus in the two
bulks coincides with the $\Delta V$ discussed above: its value for
this calculation is 2.74 eV. Two independent self-consistent
calculations for the individual bulks are then performed: we find that
the electrostatic-potential average is 8.65 eV below the Fermi level
in bulk Al, while it is 5.17 eV below the valence-band top in bulk
GaAs. Putting these three figures together, we get the value
$\Phi_{p}$=0.74 eV for the Schottky barrier at our ideal junction
between GaAs and fake Al. When we compare different (001) calculations
amongst themselves, as extensively done below, our estimated numerical
accuracy for $\Phi_{p}$ is 0.01 eV. We stress that this is a {\it
relative} accuracy, for a given set of technical
ingredients. Variation of the latter, as for instance by adopting
different pseudopotentials, would affect the results by much more.
We now investigate how our calculated value of $\Phi_{p}$ depends on
different perturbations which alter the interface morphology. 

First of all we insert a thick layer of vacuum between the metal and
the semiconductor: the calculated value of the barrier becomes thus
equal to the difference between the work functions of the metal and of
the semiconductor. Technically, we perform the calculation in the same
geometry as in Fig.~\ref{base}, but {\it removing} the Ga$_1$ and
As$_1$ layers. We find in this way a barrier of $-0.24~{\rm eV}$, very
much different from the previously calculated value of
$\Phi_{p}=0.74~{\rm eV}$. This result provides further evidence (if
any was needed) that the early Mott-Schottky model---where the
identity of the two quantities was postulated---is invalid. 

We consider then a very {\it thin} layer of vacuum: instead of
breaking the Al--As bond, we gently elongate it while keeping the rest
of the structure rigid (the length of the supercell is elongated
accordingly). Such a displacement is commonly referred to as {\it
interfacial strain}. The Schottky barrier is found to depend very
weakly upon interfacial strain: it takes in fact a strain as large as
3\%\ in order to vary $\Phi_{p}$ by 0.01 eV, our estimated numerical
accuracy.  With the (enormous) value of 10\%, $\Phi_{p}$ varies by
about 0.04 eV. 

Next we perform an analogous 10\%\ elongation, but on the Ga--As bond
nearest to the interface: we get in this case the much larger
variation of 0.09 eV. We give below a simple rationale for such
different dependence of $\Phi_{p}$ on different local strains: we will
see that the effective charges of interface ions play a major role.

The next step is to consider the effect of {\it bulk} strain on the
metal side.  Of course in the epitaxial geometry only uniaxial
tetragonal strain is allowed, where the Al lattice constant along the
growth axis is elongated by a factor $1\!+\!\epsilon$. The calculated
$\Phi_{p}$ is completely insensitive to $\epsilon$: a calculation
performed for $\epsilon\!=\!0.01$ gives a $\Phi_{p}$ variation of 0.01
eV. The $\epsilon$ value of 1\%\ corresponds to the actual
mismatch-induced relaxation of an epitaxial Al slab (when we choose
the Al bulk equilibrium lattice constant equal to the theoretical
one). This finding is rather unexpected, since---according to previous
theoretical work---the barrier for a given semiconductor seems to vary
with the nature of the metal.\cite{vanSchilfgaarde90} Instead we find
that the barrier is unchanged in the special case considered, namely
two metals having the same chemical composition but different lattice
parameters, hence different electronic densities.

We elaborate a little bit more about these findings, which give
insight into the robustness of $\Phi_{p}$ and shed some light on the
very important---although disturbingly vague---concept that the
barrier is formed extremely close to the semiconductor.\cite{Monch}
Imagine an ideal double interface, where the semiconductor is joined
to a first metal, and then the first metal is joined to a second
metal. The barrier forms at the semiconductor/metal interface, and
then---if the middle slab is thick enough---remains constant through
the second interface, since the Fermi level is aligned across any
metal/metal contact. This {\it transitivity rule} is not expected to
hold when the thickness of the middle slab is reduced. Instead, in our
case study a macroscopic slab is not needed---not even a microscopic
one---in order for the barrier to be robustly established. As a double
check of our transitivity finding, we scrutinize the two contributions
$\Delta V$ and $\Delta E_p$ separately: while their sum turns out to
be $\epsilon$-independent, their individual variation is
sizeable. With the above value of $\epsilon\!=\!0.01$, the calculated
$\Delta E_p$ varies by $-0.10$ eV: we wish to compare this to the
$\Delta V$ value at an ideal strained/unstrained metal
homojunction. To this aim, a supercell calculation is unnecessary:
$\Delta V$ is a pure volume effect, and we get it by calculating the
deformation potential\cite{rap68} of bulk Al, {\it i.e.} the linear
variation of the Fermi energy, measured with respect to the average of
the electrostatic potential. We find in this way $\Delta V = -0.11$
eV, in very good agreement with the above value.

The next probe we are going to use in order to test the robustness of
the barrier height, are displacements of individual atoms, while the
rest of the structure is kept fixed. The basic quantities measuring
the response of the electronic system to such perturbations are the
effective charges for lattice dynamics. Consider a displacement of an
ionic plane in the bulk semiconductor by an amount $u$: this creates a
dipole per unit area, inducing a potential lineup of $\Delta V = 8\pi
e^2 Z^*_{\rm T} u / (\varepsilon_\infty a^2)$, where $a$ is the cubic
lattice constant, $\varepsilon_\infty$ is the dielectric constant, and
$Z^*_{\rm T}$ is the Born (alias transverse) effective charge of the
given ionic species.\cite{Kunc85} Given the composite nature of our
heterostructure, it proves better to deal with $Z^*_{\rm T}$ and
$\varepsilon_\infty$ altogether: we focus then on the {\it
longitudinal} effective charges $Z^*\! =\! Z^*_{\rm T} /
\varepsilon_\infty$. The bulk GaAs value appropriate to our
computational framework is $Z^* \!=\! \pm 0.18$, while in any bulk
metal $Z^*$ vanishes due to perfect screening.\cite{nota} The
calculation of the effective charges of the different ions across the
junction gives a way to monitor the transition between the two bulk
materials and provides a very meaningful measure of the interface
thickness. In fact, a structural distortion may affect (to linear
order) the electrostatic lineup---and hence the barrier
$\Phi_p$---only if it displaces ions whose $Z^*$ is nonvanishing.

\begin{figure}  \centerline{\psfig{file=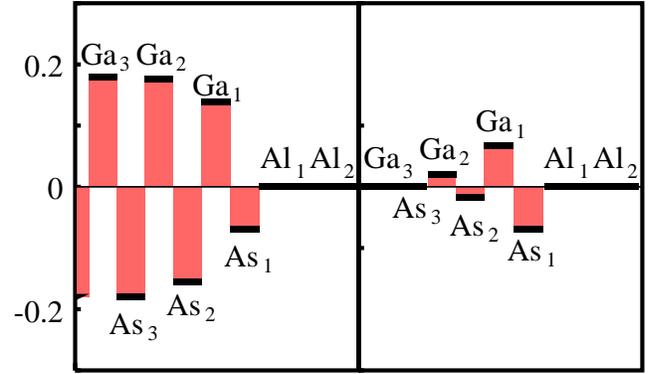,height=6cm}}
\caption[1]{The left panel shows the calculated dynamical
charges in the form of an hystogram; the darkest regions indicate our
numerical accuracy in the calculation. The right panel is the macroscopic
average of the left one: it shows the averages of the $Z^*$'s over a
segment, centered at a running point, and whose length equals the
periodicity of the bulk semiconductor region. The plot illustrates the
dynamical neutrality of the interface, and also shows that the interface
region is more extended on the semiconductor side than on the metal one. }
\label{histogram} \end{figure}

Our calculations follow Ref.~\onlinecite{Kunc85}, with a tipical $u$
value of 0.03 a.u.; a conservative estimate\cite{nota} of the
numerical accuracy of our $Z^*$'s is 0.01. When approaching the
interface from the semiconductor side, our calculated $Z^*$ values
are: $+0.18$ (Ga$_2$), $-0.15$ (As$_2$), $+0.14$ (Ga$_1$), $-0.07$
(As$_1$). Entering into the metal, the calculated $Z^*$ drop rapidly
to their (vanishing) bulk value. Since there are two nonequivalent Al
atoms per plane, we displace each of them at a time. We get $\pm 0.01$
for Al$_1$, and $-0.01$, $+0.02$ for the Al$_2$ atoms. These figures 
(also shown in Fig.~\ref{histogram}) have been
rounded to 0.01: their apparent differences being of the order of our
numerical accuracy. One important message emerging from our calculated
$Z^*$'s is that---as far as the effective charges are concerned---the
interface is very sharp on the metal side, while instead a
semiconductor ion ``feels'' the presence of the metal up to a depth of
a few layers: the closest cation (Ga$_1$) is already strongly
``nonmetallic'', though not yet bulklike. Although $Z^*$ is a {\it
linear} property of lattice distortions, our calculations indicate
that a structural defect on the metal side---even very close to the
junction---would have a little effect on $\Phi_p$; while on the
contrary a defect on the semiconductor side is likely to have a
sizeable effect.  Of particular importance to the barrier height are
therefore the detailed arrangements of the semiconductor atoms closest
to the metal (given that noncentrosymmetric structural defects deep in
the semiconductor can be ruled out). This sensitivity of the barrier
height to the morphology of the first few semiconductor layers is in
qualitative agreement with the findings of other
authors,\cite{vanSchilfgaarde90,vanSchilfgaarde94,Berthot96} who have  
considered chemical defects in an otherwise undistorted structure.

We have recently discovered a novel sum rule for the dynamical charges
at the surface of a semi-infinite crystal,\cite{rap94} which is easily
generalized to the case of an interface between a pair of semiinfinite
crystals.  The present (001) geometry is a particularly simple
example, where the meaning of our sum rule can be made clear without
any formal derivations. We first observe that the usual acoustic sum
rule\cite{PCM} (ASR) requires the sum of all $Z^*$ in the supercell to
vanish: in fact, our calculations comply with ASR within a few times
0.01. The sum rule can be interpreted as a ``dynamical neutrality'' of
the supercell as a whole: since our supercell contains two equivalent
interfaces, the ASR obviously implies the dynamical neutrality of each
of them separately. We may assume each of the interface regions to be
one half of the supercell, and clearly the sum of the $Z^*$ vanish in
each of them. The key point is that our semiconductor slab has $n$
cations and $n+1$ anions ($n=6$ in the actual calculation), and
therefore the central anion must be reckoned with weight {\it one
half} in summing the dynamical charges of each interface. One arrrives
therefore at the important conclusion---which applies in general to
any {\it isolated} (001) metal/semiconductor interface---that the sum
of the dynamical charges $Z^*$ in the interface region equals {\it one
half} the bulk dynamical charge of the semiconductor (with the
appropriate sign). As a corollary, the semiconductor ions in the
interface region {\it cannot} have the same dynamical charges as in
the bulk. All this is in perfect agreement with our computational
findings.

The effective charges are very closely related to the lineup induced
(to linear order) by interface strain, as first shown in
Ref.~\onlinecite{rap64} for the similar case of a
semiconductor-semiconductor heterojunction. In the present case we
have independently calculated the effect of interface strain (see
above) and found that it is very small. More precisely, we find zero
$\Phi_p$ variation (within our computational tolerance) when the
bond-length elongation is comparable to the one used in calculating
the $Z^*$'s. The explanation for this finding lies in the fact that
all the effective charges on the metal side are extremely small. Let
us think of an isolated junction between two semiinfinite bulks: the
interface strain amounts to a rigid relative translation. Suppose
first that the semiconductor is kept fixed, and that the metal is
displaced: by linearity, the lineup induced by the displacement of the
semiinfinite metal is the sum of the lineups induced by the
displacement of individual metal planes, and this sum is close to zero
using our calculated $Z^*$ values. We wish to recover an identical
result when we keep the metal fixed, and we displace the semiconductor
instead: this looks less trivial, since the effective charges
oscillate indefinitely in the semiconductor bulk.  We have shown in
Ref.~\onlinecite{rap94} how to regolarize such an indeterminate sum
using the appropriate physical criterion: to the present purposes,
suffices to say that the dynamical neutrality of the interface,
discussed above, is the crucial property ensuring the correct result.

In conclusion, we have shown that the effective dynamical charges
$Z^*$ in the interface region are the key quantity for rationalizing
morphology-induced variations of the Schottky barrier. A detailed
study of these charges show which distortions affect (or do not
affect) the barrier height. Actual calculations performed for
As-terminated Al/GaAs(001) show that the semiconductor $Z^*$ converge
to their bulk value rather slowly: the actual thickness of the
interface region, when monitored by means of $Z^*$, is definitely
larger than an analysis of the mere static electronic charge would
suggest. Finally, owing to a novel sum rule,\cite{rap94} the sum of
all $Z^*$ in the interface region equals one half the bulk $Z^*$
value.

We thank S. de Gironcoli for several illuminating discussions and for
invaluable technical help. We also thank the authors of
Ref.~\onlinecite{Bardi96} for providing a copy of their paper before
publication and for useful discussions.


\begin{references}
\bibitem{nota} Bulk $Z^*$ calculation have been performed using planar
displacements in pure Al (or pure GaAs) supercells, and ingredients analogous
to those used elsewhere in this work. For Al we get $|Z^*|$ smaller than
0.005; for GaAs, we agree with the experiment within 10\%.
\end{references}
\end{document}